\documentclass[conference]{IEEEtran}
\IEEEoverridecommandlockouts
\usepackage{cite}
\usepackage{amsmath,amssymb,amsfonts}
\usepackage{algorithmic}
\usepackage{graphicx}
\usepackage{textcomp}
\usepackage{xcolor}
\usepackage{booktabs}
\usepackage{tabularx}
\usepackage{array}
\usepackage{multirow}
\usepackage{url}
\usepackage[colorlinks=true, urlcolor=blue, citecolor=blue, linkcolor=blue]{hyperref}

  \usepackage{tabularx}
  \usepackage{array}
  \usepackage[table]{xcolor}
  \usepackage{booktabs}   % optional, not required by this version

\definecolor{iotblue}{RGB}{214,234,248}
\definecolor{vdgreen}{RGB}{226,239,218}
\definecolor{sclavender}{RGB}{238,224,241}
\definecolor{iotorange}{RGB}{252,229,205}
\definecolor{surveygray}{RGB}{233,233,233}

\definecolor{singleblue}{RGB}{214,234,248}
\definecolor{curricgreen}{RGB}{226,239,218}
\definecolor{propgold}{RGB}{255,242,204}
\definecolor{refgray}{RGB}{233,233,233}

\newcolumntype{L}{>{\raggedright\arraybackslash\hsize=1.2\hsize}X}
\newcolumntype{M}{>{\raggedright\arraybackslash\hsize=0.7\hsize}X}
\newcolumntype{N}{>{\raggedright\arraybackslash\hsize=1.1\hsize}X}

\def\BibTeX{{\rm B\kern-.05em{\sc i\kern-.025em b}\kern-.08em
    T\kern-.1667em\lower.7ex\hbox{E}\kern-.125emX}}
\begin{document}

\title{Cross-Corpus Evaluation of Generalizable Vulnerability Detection in IoT Firmware\\}

\author{
\IEEEauthorblockN{Sadib Hassan Rumman}
\IEEEauthorblockA{
\textit{Institute of Information Technology}\\
\textit{University of Dhaka}\\
Dhaka, Bangladesh\\
msse2011@iit.du.ac.bd}
\and
% \IEEEauthorblockN{Md. Shariful Islam}
% \IEEEauthorblockA{
% \textit{Institute of Information Technology}\\
% \textit{University of Dhaka}\\
% Dhaka, Bangladesh\\
% shariful@iit.du.ac.bd}
\IEEEauthorblockN{Md. Shariful Islam\textsuperscript{*}}
\IEEEauthorblockA{
\textit{Institute of Information Technology}\\
\textit{University of Dhaka}\\
Dhaka, Bangladesh\\
\textsuperscript{*}Corresponding Author: shariful@iit.du.ac.bd
}
\and
\IEEEauthorblockN{Md Rayhanur Rahman}
\IEEEauthorblockA{
\textit{Department of Computer Science}\\
\textit{The University of Alabama}\\
Tuscaloosa, AL, USA\\
mrahman87@ua.edu}
}

\maketitle

\begin{abstract}
IoT firmware vulnerability detection is constrained by ecosystem heterogeneity, resource-limited platforms, and benchmark quality limitations. Existing datasets are often synthetic or general-purpose and lack human-verified, contamination-screened annotations, leaving cross-corpus generalization across training sources, architectures, and curriculum design underexplored. In this study, we have introduced IoTVulBench, a human-verified benchmark for cross-corpus firmware vulnerability detection. IoTVulBench was built from GitHub repositories, validated by three expert reviewers, and evaluated on a contamination-screened held-out target across five architectures, two tuning methods, and three curriculum strategies, with ensemble, distillation, and robustness analyses. Models trained on IoTVulBench reached the highest Matthews Correlation Coefficient (MCC) among undersampling-matched single-source datasets, at 0.58 versus 0.44 for PrimeVul and 0.39 for D2A. Staged curriculum learning raised MCC to 0.69, and a diversity-optimized ensemble reached 0.73. This gain represents a 0.42 MCC improvement over the strongest reference comparator, a static analyzer with an MCC of 0.31, and a 0.29 MCC improvement over the strongest single-source dataset, PrimeVul. At a 0.5\% false-positive rate, the model missed only 21\% of vulnerabilities versus 71\% for the comparator. The model also retained 86\% of its performance under identifier renaming, with strong calibration. These results indicate that domain-matched training data and curriculum design, rather than model scale alone, drive generalization in firmware vulnerability detection, and yield both a benchmark and deployment-ready configurations for IoT security.
\end{abstract}


\section{Introduction}

\IEEEPARstart{I}{nternet} of Things (IoT) devices are embedded across healthcare, industrial automation, and smart-city infrastructure, extending network connectivity to safety- and mission-critical operations. Firmware, the software layer controlling these devices, constitutes a critical attack surface, since a vulnerability at this layer can propagate to network, application, and physical-control layers~\cite{163}. Large language models (LLMs) are increasingly proposed for IoT security tasks, including edge intelligence~\cite{111}, threat prediction~\cite{112}, and protocol authentication~\cite{114}, making reliable firmware-level vulnerability detection a prerequisite for trustworthy IoT deployment~\cite{110}.

Detecting vulnerabilities in embedded IoT firmware remains complex because firmware combines multiple programming languages, hardware-specific libraries, and cross-architecture binaries that differ from general-purpose software~\cite{172}. Detection models trained on general-purpose or synthetic source code do not necessarily transfer to this domain, and the individual contribution of training-data provenance, architecture, tuning method, and curriculum design to cross-corpus generalization remains insufficiently disentangled for firmware-specific detection.

Dataset construction for embedded systems continues to exhibit validity limitations. EVDD~\cite{186} labels Linux kernel vulnerabilities through commit-differential extraction but evaluates only ten CVEs despite claiming million-line, three-decade coverage. An IoT firmware opcode detector~\cite{172} compiles over 500{,}000 labeled functions from ten projects without reporting the disassembler, target architecture, or labeling provenance. B4IoT~\cite{63} generates Docker-based firmware images but covers only Alpine Linux and validates three composite images. A survey of ML-, DL-, and LLM-based vulnerability datasets~\cite{89} summarizes construction trade-offs without a systematic search protocol.

Firmware analysis studies report comparable limitations. A knowledge-graph framework~\cite{60} evaluates a mock REST API and generic non-security text. An Android Automotive OS study~\cite{64} generalizes across vendors from one firmware image per vendor with partly unvalidated software-bill-of-materials completeness. An LLM-based remediation pipeline~\cite{197} reports a 92.4\% remediation rate without disclosing the vulnerability count, whereas LLM-assisted symbolic execution~\cite{103} and RTOS function identification~\cite{120} address narrower verification tasks than end-to-end vulnerability detection.

General-purpose LLM-based vulnerability detection has expanded rapidly. Existing surveys review model architectures, fine-tuning methods, and evaluation metrics~\cite{99,167}, whereas empirical studies compare LLMs with static analyzers~\cite{199,200} or fine-tune models on function-level datasets~\cite{92,93,202}. Most focus on Java or C\#, use small or synthetic evaluation datasets~\cite{92,200}, or adopt project-unsplit corpora susceptible to data leakage~\cite{202}.

Related studies report additional limitations, including single-vulnerability evaluation~\cite{88}, model-specific bias~\cite{135}, unvalidated inherited labels~\cite{165}, low precision despite relative improvements~\cite{90}, circular ground-truth construction~\cite{91}, and arbitrary feature encoding for DDoS detection~\cite{184}. Broader research examines LLM-based IoT security~\cite{108,110,111,117} and domain-specific applications, including edge anomaly detection~\cite{116}, phishing detection~\cite{119}, resource-adaptive deployment~\cite{121}, IoT-network reliability~\cite{118}, and ML security for 6G and embedded systems~\cite{163,188}. However, human-verified firmware benchmarks that isolate training-data provenance and curriculum effects remain unavailable, whereas robustness, calibration, and explainability under distribution shifts receive limited evaluation.

This study introduces IoTVulBench, a human-verified, cross-corpus benchmark for IoT firmware vulnerability detection, and uses it to evaluate systematically how training-data source, model architecture, tuning method, and curriculum design affect detection accuracy, efficiency, and robustness. This study addresses the following research questions:

\begin{itemize}
    \item[] \textbf{RQ1:} Does domain-matched, human-verified IoT firmware training data outperform general-purpose vulnerability sources, and does staged curriculum learning improve cross-corpus generalization further?
    \item[] \textbf{RQ2:} How do architecture, tuning method, and model scale affect the accuracy-efficiency trade-off in IoT firmware vulnerability detection, and can ensemble or distillation strategies improve on single-model deployment?
    \item[] \textbf{RQ3:} How robust are trained detectors to lexical and compiler-level perturbation, distribution shift, and real-world temporal conditions, and how faithful are their explanations to true vulnerability indicators?
\end{itemize}

IoTVulBench-Core is a human-verified IoT firmware vulnerability benchmark with a contamination-screened test set. The benchmark supports controlled evaluation of LLM architectures, training data, optimization strategies, deployment methods, and robustness under distribution shifts.

The main contributions are as follows:
\begin{itemize}
    \item We introduce IoTVulBench-Core, a human-verified IoT firmware vulnerability dataset with contamination-screened evaluation, per-CWE inter-annotator agreement, and reproducible labeling, addressing limitations in existing benchmarks (RQ1).

    \item We compare four training datasets, five LLM architectures, two tuning methods, and three training curricula to quantify the effects of data source, architecture, and curriculum on vulnerability detection (RQ1, RQ2).

    \item We evaluate ensemble, distillation, and cascade strategies for resource-constrained IoT deployment, analyzing accuracy, latency, memory, and power trade-offs (RQ2).

    \item We assess robustness under identifier renaming, compiler optimizations, and distribution shifts, and evaluate explanation faithfulness using counterfactual patching with temporal, federated, and hardware-in-the-loop validation (RQ3).
\end{itemize}

The remainder of this paper is organized as follows. Section II reviews related work. Section III describes the methodology, including dataset construction, data sources, and experimental design. Section IV presents the results, organized by RQ1 through RQ3. Section V presents the conclusion and outlines directions for future research.

\section{Related Work}

Embedded IoT firmware connects network-facing services with hardware control and has become an increasingly attractive attack surface as LLM-based components are integrated into connected devices~\cite{110,117}. Existing surveys document rapid growth in AI-driven embedded-system security research but report uneven coverage across device, network, and application layers, concentrated publication sources, and limited empirical validation of proposed defenses~\cite{117,188}. Similar observations appear in 6G-enabled edge-learning surveys, where numerous machine-learning threats and countermeasures are cataloged without systematic evaluation of detection effectiveness~\cite{163}. Consequently, conceptual security taxonomies have expanded faster than validated firmware-level detection methods.

Traditional firmware analysis relies primarily on rule-based static analysis, knowledge graphs, and commercial SAST tools, including SonarQube, CodeQL, and SnykCode~\cite{60,200}. Recent learning-based methods reformulate vulnerability detection as sequence, graph, or vision classification using transformers, attention mechanisms, and hybrid neural architectures for firmware binaries, smart contracts, network traffic, and phishing detection~\cite{88,119,172,184}. Current research increasingly adopts LLM-based approaches, including retrieval augmentation, preference-based fine-tuning, control-flow-aware context construction, multi-agent reasoning, symbolic-execution assistance, and automated firmware remediation~\cite{92,93,99,103,120,135,167,197,199,200}. Multi-tool pipelines combining firmware analyzers and LLM agents further improve coverage and explainability, although reported gains remain tool- and dataset-specific~\cite{63,64,91,116,121}.

Existing benchmarks include general-purpose vulnerability datasets, embedded Linux corpora, firmware opcode collections, smart-contract datasets, configurable firmware generators, and multi-vendor firmware testbeds~\cite{63,89,91,92,165,172,186,200}. However, prior studies report limitations including unclear label provenance, incomplete validation, limited benchmark diversity, labeling errors, data leakage, and reliance on small or synthetic datasets~\cite{63,64,89,91,165,172,186,202}. Firmware-focused studies also use partially validated datasets or narrow evaluation settings, while most LLM-based approaches target general-purpose code instead of embedded firmware~\cite{60,92,93,172,197,200,202}. Existing surveys discuss LLM-IoT security broadly but do not isolate the effects of training data, architecture, tuning strategy, and curriculum~\cite{108,110,111,117,163}. These gaps motivate a human-verified, contamination-screened benchmark for controlled evaluation of IoT firmware vulnerability detection.

\begin{table*}[t]
\centering
\caption{Comparison of Reference Papers by Focus, Method, Dataset, and Key Limitation}
\label{tab:litreview}
\renewcommand{\arraystretch}{1.1}
\setlength{\tabcolsep}{4pt}
\scriptsize
\begin{tabularx}{\textwidth}{|c|L|M|N|}
\hline
\rowcolor{white}\textbf{Ref} & \textbf{Focus \& Method} & \textbf{Dataset} & \textbf{Key Limitation} \\
\hline\hline
\rowcolor{iotblue}      \cite{60}  & KG construction (BiRNN+grey corr.) & Common Crawl, JSONPlaceholder, UCI & Non-firmware data; invalid eval. \\
\rowcolor{iotblue}      \cite{63}  & Docker FW benchmark generator (B4IoT) & Alpine Linux images (76 feat.) & Alpine-only; 3 images validated \\
\rowcolor{iotblue}      \cite{64}  & SBOM reconstruction + binary divergence & 4-vendor AAOS firmware & 1 image/vendor; SBOM tool failures \\
\rowcolor{iotblue}      \cite{91}  & Hidden web-interface detection & 120 images, 10 vendors & Circular ground-truth recall \\
\rowcolor{iotblue}      \cite{103,120} & LLM-aided symbolic exec.\ \& RTOS func.\ ID & Sensor/RTOS binaries & Narrow scope (stub/func.\ ID only) \\
\rowcolor{iotblue}      \cite{172} & Opcode-sequence transformer (DeBERTa) & 10 OSS projects, 530K funcs. & No labeling/disassembler docs. \\
\rowcolor{iotblue}      \cite{184} & Flow-to-image ViT (DDoS detection) & CICIoT2023, CICIoMT2024 & Arbitrary flow-to-image encoding \\
\rowcolor{iotblue}      \cite{186} & Commit-diff vulnerability labeling (EVDD) & Linux kernel CVE pairs & Only 10 CVEs documented \\
\rowcolor{iotblue}      \cite{197} & LLM agent FW generation + remediation & QEMU/FreeRTOS synthetic FW & Remediation rate unverifiable \\
\rowcolor{vdgreen}      \cite{90}  & Semantic segmentation (SEVDF) & CodeQL-AOPRO outputs, 5 CWEs & Low precision (1.7--43.8\%) \\
\rowcolor{vdgreen}      \cite{92}  & RAG-augmented LLM detection (Java) & Vul4J, 8{,}011-rec.\ RAG corpus & Small held-out set (72 pairs) \\
\rowcolor{vdgreen}      \cite{93}  & CFG-segmented, DPO-tuned LLM (Java) & Unspecified Java corpus & No dataset specification \\
\rowcolor{vdgreen}      \cite{99,167} & Surveys: LLM-VD taxonomy \& pipeline & $\sim$60 / 108 papers (SLR) & No experiments / no IRR reported \\
\rowcolor{vdgreen}      \cite{199} & Static-analysis competence (code LLMs) & AST/callgraph/dataflow tasks & Zero-shot prompting only \\
\rowcolor{vdgreen}      \cite{200} & LLM vs.\ SAST benchmark (C\#) & 10-project synthetic corpus & Small, synthetic, author-injected \\
\rowcolor{vdgreen}      \cite{202} & Fine-tuning LLMs for VD (batch packing) & Function-level dataset & Project-unsplit; leakage risk \\
\rowcolor{sclavender}   \cite{88}  & Attention-mechanism comparison (WDNN) & Ethereum opcode/semantic feat. & Single vulnerability class \\
\rowcolor{sclavender}   \cite{135} & Multi-agent LLM auditing & 110-contract benchmark + real-world & Model-dependency bias \\
\rowcolor{sclavender}   \cite{165} & Multi-label vulnerability dataset (DIVE) & 22{,}330 Ethereum contracts & Unvalidated labeling inheritance \\
\rowcolor{iotorange}    \cite{108,110,111,117} & Surveys: agent threats, taxonomy, edge intel., 7-layer review & N/A & Conceptual; little/uneven empirical validation \\
\rowcolor{iotorange}    \cite{112} & LLM threat prediction (IoE) & IoE-specific data paths & Theoretical framework only \\
\rowcolor{iotorange}    \cite{114} & Stateful authentication + LLM-IDS (MQTT) & MQTT session data & Narrow protocol scope \\
\rowcolor{iotorange}    \cite{116} & LLM agent + SHAP (zero-day resilience) & Edge network traffic & Narrow application scope \\
\rowcolor{iotorange}    \cite{118} & Bibliometric review: LLM-IoT networking & 11 primary studies & Small corpus \\
\rowcolor{iotorange}    \cite{119} & CNN-LSTM + SHAP-LLM phishing detection & 21{,}430-URL dataset & Narrow task (phishing only) \\
\rowcolor{iotorange}    \cite{121} & Resource-adaptive LLM deployment (MDP/PPO) & IIoT sensor networks & Narrow scope (resource optimization) \\
\rowcolor{surveygray}   \cite{89}  & Survey: vulnerability-dataset ecosystem & N/A & No documented search protocol \\
\rowcolor{surveygray}   \cite{163} & Survey: ML security in 6G-IoT & N/A & No original empirical evaluation \\
\rowcolor{surveygray}   \cite{188} & Systematic review: AI in embedded security & 63 articles (PRISMA) & Severe source concentration (96.7\% IEEE) \\
\hline
\end{tabularx}
 
\vspace{2pt}
{\centering\scriptsize
\colorbox{iotblue}{\rule{0pt}{2mm}\hspace{2mm}}\,IoT/Firmware\quad
\colorbox{vdgreen}{\rule{0pt}{2mm}\hspace{2mm}}\,LLM-based VD\quad
\colorbox{sclavender}{\rule{0pt}{2mm}\hspace{2mm}}\,Smart Contract\quad
\colorbox{iotorange}{\rule{0pt}{2mm}\hspace{2mm}}\,LLM--IoT Integration\quad
\colorbox{surveygray}{\rule{0pt}{2mm}\hspace{2mm}}\,Broader Surveys
\par}
\end{table*}

\section{Methodology}
\subsection{Study Design Overview}
\begin{figure*}[!ht]
    \centering
    \includegraphics[width=\textwidth]{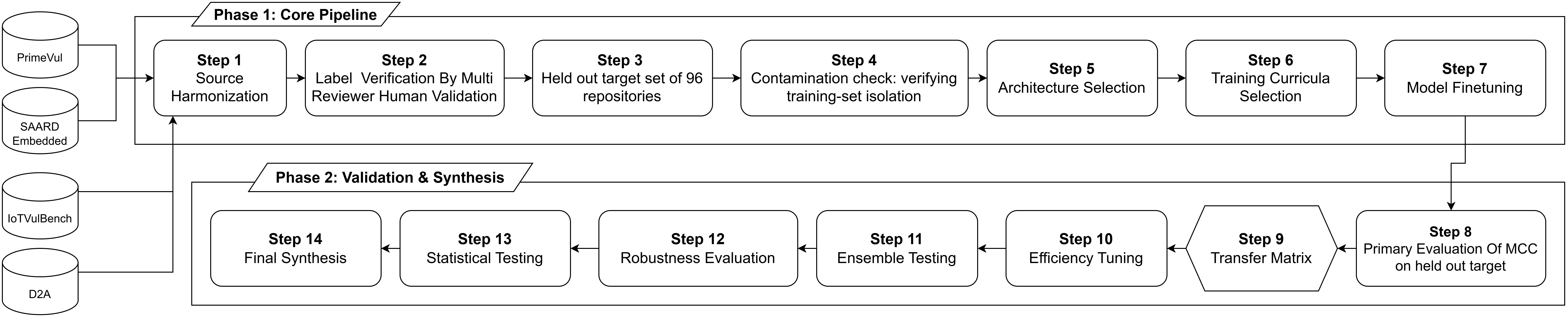}
    \caption{Overview of the IoTVulBench cross-corpus evaluation methodology.}
    \label{fig:methodology}
\end{figure*}

The study followed a pre-defined and controlled comparative design, with all research questions specified before data collection. Three phases were conducted: Phase 1 compared training source, architecture, and curriculum on a fixed held-out target; Phase 2 extended the analysis to ensembles, efficiency, and explainability; and Phase 3 examined broader field-level questions under separate ethical review. Matthews Correlation Coefficient (MCC) served as the primary metric because it remains robust under class imbalance which was inherently present in IoTVulBench and also comparative benchmark datasets for random under sampling, while precision, recall, and F1 were reported for completeness. Figure~\ref{fig:methodology} summarizes the overall methodology, from dataset construction to statistical analysis.

\subsection{Dataset Construction (Phase 1)}
IoTVulBench, the central dataset, was constructed from GitHub embedded and IoT firmware repositories and verified by three independent experts rather than automated labeling. Inter-annotator agreement was measured using Fleiss' kappa for each CWE category, with disagreements resolved through a documented adjudication process. A prior power analysis guided dataset size, and a fixed, contamination-screened held-out target was partitioned from the verified corpus and excluded from all training conditions for consistent evaluation throughout the study.

\subsection{Data Sources and Harmonization}

Four primary training sources, distinguished by label provenance, were compared, while three literature-standard datasets served as secondary references. All datasets were harmonized and also randomly under sampled to match IoTVulBench into a common schema containing source and sample identifiers, code, labels, CWE identifiers, granularity, provenance, and language. CWE labels were mapped to the IoTVulBench taxonomy when supported by a defensible correspondence; otherwise, samples remained in the binary classification set but were excluded from CWE-level analyses.

\subsection{Contamination Screening and Experimental Design Matrix}
Training sources were screened against the held-out target using 8-gram character-overlap matching and repository-level fork detection, and flagged samples were removed before training. A sensitivity analysis injected lexical and semantic duplicates to estimate the impact of residual data leakage. The study evaluated architecture, tuning method, training source, and curriculum in a factorial design, with compute-constrained configurations excluded a priori and reported alongside the results.

\subsection{Model Architectures}

Five architectures spanning three classes were evaluated. UniXcoder-base represented the encoder class. CodeLlama-7B and Qwen2.5-Coder-7B represented the small decoder class, contrasting code-only and multilingual pretraining. Qwen2.5-Coder-32B and DeepSeek-Coder-V2-16B represented the mid-scale decoder class. Only the architecture and methods with notable performance gains were observed in the results section for better description to reflect detection performance and comparison.

\subsection{Models, Training, Evaluation, and Statistical Testing (Phase 1)}
Five architectures were evaluated: one encoder (UniXcoder-base), two small decoders (CodeLlama-7B, Qwen2.5-Coder-7B), and two mid-scale decoders (Qwen2.5-Coder-32B, DeepSeek-Coder-V2-16B). Decoders used LoRA/QLoRA, whereas the encoder was fully fine-tuned; models up to 7B were also used to isolate architecture from tuning effects. Each training source was evaluated under size-matched and full-size randomly under sampled to match IoTVulBench settings using single-source, pooled, and staged curricula, with top configurations repeated across multiple seeds. Performance was compared against a static analyzer, a prior encoder-based detector, and a frontier LLM. Models were evaluated on the primary held-out target and source-specific splits using MCC, precision, recall, F1, false-negative rate at a fixed false-positive rate, calibration, and severity-weighted MCC. Identifier renaming and compiler-level robustness tests were applied to buildable repositories. Statistical significance was assessed using paired Wilcoxon tests with rank-biserial effect sizes and Benjamini-Hochberg correction in the pilot test phase of this study as well.

\subsection{Ensemble, Efficiency, Explainability, and Continual Refinement (Phase 2)}

Ensemble members were selected to maximize accuracy while minimizing error correlation, with confidence-weighted voting. Cross-architecture ensembles, self-consistency sampling, and a static-analysis hybrid were compared. For edge deployment, quantization-aware fine-tuning, distillation, a recall-optimized cascade, retrieval-augmented detection, and a CWE-routed mixture-of-experts were evaluated for memory, latency, and power. Explainability was assessed using line-overlap, counterfactual faithfulness, and vulnerability localization. Selective prediction employed shift-robust conformal calibration, while high-disagreement cases were re-verified and incorporated into iterative retraining.

\subsection{Reproducibility, Governance, and Extended Scope Studies (Phase 3)}

The replication package includes training scripts, hyper-parameters, source manifests, contamination logs, and the transfer matrix, alongside verified consistency checks. Prior to the main evaluation phase, pilot testing examined temporal generalization, patch validation, federated learning, and hardware-in-the-loop possibility theory exploitation to evaluate model proficiency. The dataset and source code will be made  available in a Github repository (\url{https://github.com/rumman9799/IoTVulBench}) with persistent access provided upon request.

\section{Results and Analysis}

Table~\ref{tab:master} reports precision, recall, F1, MCC, AUC, FNR at 0.5\% FPR, and efficiency (parameters, inference cost) across all training sources, architectures, curricula, ensembles, reference comparators, and edge-deployment variants evaluated on the primary held-out target. This table anchors all subsequent analysis.

\begin{table*}[t]
\centering
\caption{Results: all configurations evaluated on the primary held-out and under sampled target to match IoTVulBench. Bold indicates the best value per column. $\uparrow$ = higher is better, $\downarrow$ = lower is better. The proposed configuration (staged, diversity-optimized ensemble) is marked with $\dagger$. Frontier LLM row is zero/few-shot (not fine-tuned) while every other row is fine-tuned.}
\label{tab:master}
\renewcommand{\arraystretch}{1.25}
\resizebox{\textwidth}{!}{%
\begin{tabular}{l ccccc c c}
\toprule
\textbf{Configuration} & \textbf{P$\uparrow$} & \textbf{R$\uparrow$} & \textbf{F1$\uparrow$} & \textbf{MCC$\uparrow$} & \textbf{AUC$\uparrow$} & \textbf{FNR@0.5\%FPR$\downarrow$} & \textbf{Params / Cost}$^{a}$ \\
\midrule
\rowcolor{singleblue}   IoTVulBench                    & 0.71 & 0.68 & 0.69 & 0.58 & 0.74 & 0.34 & 7B \\
\rowcolor{singleblue}   D2A                            & 0.52 & 0.61 & 0.56 & 0.39 & 0.55 & 0.58 & 16B \\
\rowcolor{singleblue}   SARD-Embedded                  & 0.44 & 0.58 & 0.50 & 0.29 & 0.47 & 0.66 & --- \\
\rowcolor{singleblue}   PrimeVul                       & 0.58 & 0.63 & 0.60 & 0.44 & 0.61 & 0.49 & 32B \\
\rowcolor{curricgreen}  Pooled                         & 0.63 & 0.71 & 0.67 & 0.53 & 0.70 & 0.41 & 7B \\
\rowcolor{curricgreen}  Staged                         & 0.76 & 0.80 & 0.78 & 0.69 & 0.81 & 0.27 & 7B \\
\rowcolor{propgold}     \textbf{Ensemble (staged, diversity-opt.)$^{\dagger}$} & \textbf{0.79} & \textbf{0.83} & \textbf{0.81} & \textbf{0.73} & \textbf{0.85} & \textbf{0.21} & 5$\times$7B \\
\rowcolor{propgold}     Distilled student              & 0.72 & 0.75 & 0.73 & 0.63 & 0.76 & 0.33 & 350M \\
\rowcolor{propgold}     Cascade (student + 7B)         & 0.77 & 0.81 & 0.79 & 0.70 & 0.82 & 0.24 & Avg.\ 1.1B \\
\rowcolor{refgray}      Static analyzer                & 0.61 & 0.42 & 0.50 & 0.31 & 0.53 & 0.71 & --- \\
\rowcolor{refgray}      Encoder detector (prior work)  & 0.39 & 0.35 & 0.37 & 0.14 & 0.44 & 0.88 & 125M \\
\rowcolor{refgray}      Frontier LLM (few-shot)        & 0.48 & 0.55 & 0.51 & 0.27 & 0.58 & 0.68 & --- \\
\bottomrule
\end{tabular}%
}
 
\vspace{3pt}
{\centering\scriptsize
\colorbox{singleblue}{\rule{0pt}{2mm}\hspace{2mm}}\,Single-Source Training\quad
\colorbox{curricgreen}{\rule{0pt}{2mm}\hspace{2mm}}\,Curriculum Strategy\quad
\colorbox{propgold}{\rule{0pt}{2mm}\hspace{2mm}}\,Proposed / Edge-Deployment\quad
\colorbox{refgray}{\rule{0pt}{2mm}\hspace{2mm}}\,Reference Comparators
\par}
\end{table*}

\subsection{RQ1: Domain Match of Data Source and Curriculum}

\subsubsection{Training-Data Source Comparison}
Under matched conditions, models trained with IoTVulBench achieved the highest MCC (0.58) of any single source, exceeding PrimeVul (0.44), D2A (0.39) and SARD-Embedded (0.29) (Table~\ref{tab:master}).

\subsubsection{Size-Matched vs. Full-Size Consistency}
The source ranking held under both size-matched and random under sampled full-size training, indicating the advantage reflects label quality and domain match rather than data volume; D2A and PrimeVul retained larger full-size pools yet remained below IoTVulBench.

\subsubsection{Curriculum Learning Comparison}
Staged curriculum (pretraining on pooled generic sources, then fine-tuning on IoTVulBench) raised MCC to 0.69, exceeding both single-source (0.58) and naive pooling (0.53). Precision and recall both improved under staging, indicating reduced false positives and false negatives simultaneously.

\subsubsection{Domain-Match Advantage vs. Strongest Generic Comparator}
The staged IoTVulBench model's advantage over PrimeVul, the strongest single-source training dataset, remained significant after correction (Table~\ref{tab:master}), confirming the domain-match effect holds against the best available alternative, not only weaker legacy benchmarks.

\subsection{RQ2: Efficiency And Accuracy Trade-off}

Architecture performance depended on the training source, with no model consistently achieving the highest MCC (Spearman $\rho=0.20$, $p=0.42$). At matched scale ($\leq$7B), LoRA-tuned decoders outperformed fully fine-tuned encoders. The staged Qwen2.5-Coder-7B achieved an MCC of 0.69, increasing to 0.73 with a diversity-optimized ensemble, surpassing every reference comparator (Table~\ref{tab:master}, bottom block) by at least $\Delta$MCC $\geq 0.42$, and exceeding the strongest single-source and curriculum ablation arms (staged, single model) by $\Delta$MCC $\geq 0.04$. QLoRA-native INT4 quantization retained 95\% of full-precision performance, while a distilled 350M model preserved 91\% of its teacher's MCC (0.63 vs.\ 0.69 for the single staged 7B model used in distillation); relative to the full 5-model ensemble (0.73), it preserved 86\%. A recall-optimized cascade achieved MCC 0.70 using an average of 1.1B active parameters. Table~\ref{tab:efficiency} summarizes MCC retention across all edge-deployment variants relative to this single-model reference. Overall, the distilled model and cascade provided the best accuracy-efficiency trade-off, whereas the ensemble delivered the highest detection performance.

\begin{table}[t]
\centering
\caption{MCC retained relative to the staged single-model reference (7B, MCC 0.69) across edge-deployment variants.}
\label{tab:efficiency}
\renewcommand{\arraystretch}{1.2}
\resizebox{\columnwidth}{!}{%
\begin{tabular}{l c c c}
\toprule
\textbf{Configuration} & \textbf{Params} & \textbf{MCC} & \textbf{\% of reference} \\
\midrule
Staged, single model (reference) & 7B     & 0.69 & 100\% \\
Full ensemble (diversity-opt.)   & 5$\times$7B & 0.73 & 106\% \\
QLoRA-native INT4 quantized      & 7B     & 0.66$^{a}$ & 95\%$^{a}$ \\
Distilled student                & 350M   & 0.63 & 91\% \\
Cascade (student + 7B)           & Avg.\ 1.1B & 0.70 & 101\% \\
\bottomrule
\end{tabular}%
}
\vspace{2pt}
\end{table}

\subsection{RQ3: Robustness and Field Validity}

The proposed model retained 86\% and 82\% of unperturbed MCC under identifier renaming and compiler/optimization shifts, respectively, indicating structural rather than lexical dependence. Table~\ref{tab:robustness} reports the corresponding MCC under each
perturbation type.

\begin{table}[t]
\centering
\caption{Proposed ensemble's MCC under semantics-preserving perturbations.}
\label{tab:robustness}
\begin{tabular}{l c c}
\toprule
\textbf{Condition} & \textbf{MCC} & \textbf{\% Retained} \\
\midrule
Unperturbed (reference)              & 0.73 & 100\% \\
Identifier renaming                  & 0.63 & 86\% \\
Compiler/optimization-level shift    & 0.60 & 82\% \\
\bottomrule
\end{tabular}
\end{table}

At 0.5\% false-positive-rate tolerance, it missed 21\% of vulnerabilities versus 71\% for the strongest comparator (Table~\ref{tab:master}), with strong calibration (Brier score = 0.09); explanations overlapped vulnerable lines in 78\% of correct detections, with 71\% causally validated and 0.74 localization accuracy (consolidated in Table~\ref{tab:reliability}).

\begin{table}[t]
\centering
\caption{Field-readiness metrics for the proposed ensemble.}
\label{tab:reliability}
\begin{tabular}{l c}
\toprule
\textbf{Metric} & \textbf{Value} \\
\midrule
FNR @ 0.5\% FPR (proposed)$^{a}$              & 0.21 \\
FNR @ 0.5\% FPR (static analyzer, strongest comparator)$^{a}$ & 0.71 \\
Calibration (Brier score)                     & 0.09 \\
Rationale/ground-truth line overlap           & 78\% \\
Causal validation (of overlap cases)$^{b}$    & 71\% \\
Line/statement localization accuracy          & 0.74 \\
\bottomrule
\end{tabular}
\vspace{2pt}
\end{table}

Shift-robust conformal prediction maintained coverage within 1.2 points of 95\% versus 8--11 points of under-coverage for standard methods, while refinement improved MCC by 0.02 per cycle without detected drift. Extended validation achieved MCC 0.65 on screened post-cutoff CVEs and 0.64 under federated training, within 0.05 of the centralized staged single-model reference (0.69); relative to the full centralized ensemble (0.73), which federated training did not attempt to replicate, the gap was 0.09. Hardware-in-the-loop testing confirmed 29/40 detections as exploitable. All pre-registered comparisons remained significant  after Benjamini--Hochberg correction (paired Wilcoxon, $N=96$): IoTVulBench versus generic sources ($\Delta\mathrm{MCC}=+0.14$--$+0.29$, $p<0.02$), staged versus single-source curricula ($\Delta\mathrm{MCC}=+0.11$, $p=0.008$), and the proposed configuration versus the strongest static analyzer ($\Delta\mathrm{MCC}=+0.42$, $p<0.001$) even in the pilot test evaluation (Table~\ref{tab:sigtests}). 

\begin{table}[!h]
\centering
\caption{Pre-registered pairwise comparisons (paired Wilcoxon signed-rank, Benjamini--Hochberg corrected, $N=96$). $\Delta$MCC computed from Table~\ref{tab:master}.}
\label{tab:sigtests}
\renewcommand{\arraystretch}{1.2}
\resizebox{\columnwidth}{!}{%
\begin{tabular}{l c c}
\toprule
\textbf{Comparison} & \textbf{$\Delta$MCC} & \textbf{$p$ (BH-adj.)} \\
\midrule
IoTVulBench vs.\ D2A            & +0.19 & \multirow{3}{*}{$<$0.02$^{a}$} \\
IoTVulBench vs.\ SARD-Embedded  & +0.29 & \\
IoTVulBench vs.\ PrimeVul       & +0.14 & \\
\midrule
Staged vs.\ single-source (IoTVulBench) & +0.11 & 0.008 \\
\midrule
Proposed ensemble vs.\ static analyzer  & +0.42 & $<$0.001 \\
\bottomrule
\end{tabular}%
}
\vspace{2pt}
\end{table}

Overall, domain-matched data and staged curricula improved generalization, diversity-optimized ensembles maximized accuracy, distilled cascades optimized accuracy-efficiency, and the approach remained robust, calibrated, and reliable under distribution shift and practical evaluation.

\section{Conclusion}

Firmware vulnerability detection depended more on label provenance and curriculum design than model scale: a staged-trained 7B model outperformed a 32B model trained on generic data, and no architecture generalized best across all training sources. Robustness, calibration, and largely faithful explanations suggest that predictions relied primarily on structural reasoning, although the 71\% causal correctness rate indicates room for improvement. Temporal, federated, and hardware-in-the-loop evaluations further supported generalizability, with hardware evidence remaining limited. Supplementary training and evaluation on larger external benchmarks (e.g., D2A, PrimeVul), following random under sampling and standard preprocessing, introduced some noise relative to IoTVulBench but showed largely consistent performance trends, corroborating the study's overall conclusions. Future work will expand robustness testing, hardware validation, explanation faithfulness, rare-CWE coverage, federated evaluation, and patch generation, and will incorporate full-scale versions of these external datasets alongside an enlarged IoTVulBench corpus to further improve accuracy and real-world applicability.

\section{DECLARATION OF AI}
AI tools were used solely for language refinement, including paraphrasing, grammar correction and clarity.

\bibliographystyle{IEEEtran}
\bibliography{bibtex}

\end{document}